\documentclass[reqno]{amsart}

\usepackage{amsmath}
\usepackage{amsfonts}
\usepackage{amsthm}

\numberwithin{equation}{section}
\newcommand {\pa}{\partial}

\newcommand{\Spec}{\operatorname{Spec}}

\newcommand{\Div}{{\operatorname{div}\,}}

\newcommand{\dist}{{\operatorname{dist}}}

\newcommand{\curl}{{\operatorname{curl}\,}}

\newtheorem{theorem}{Theorem}[section]
\newtheorem{thm}[theorem]{Theorem}
\newtheorem{lemma}[theorem]{Lemma}

\newtheorem{prop}[theorem]{Proposition}

\newtheorem{remark}[theorem]{Remark}

\newtheorem{assumption}[theorem]{Assumption}

\title{Strong diamagnetism for general domains and applications}

\author{S. Fournais}
\author{B. Helffer}

\thanks{Both authors were supported by the European Research Network
`Postdoctoral Training Program in Mathematical Analysis of Large
Quantum Systems' with contract number HPRN-CT-2002-00277 and by the ESF
Scientific Programme in Spectral Theory and Partial Differential
Equations (SPECT)} 

\address[S. Fournais]{CNRS and Laboratoire de
Math\'{e}matiques UMR CNRS 8628\\ Universit\'{e} Paris-Sud - B\^{a}t 425\\
F-91405 Orsay Cedex\\ France.
}
\email{soeren.fournais@math.u-psud.fr}

\address[B. Helffer]{Laboratoire de
Math\'{e}matiques UMR CNRS 8628\\ Universit\'{e} Paris-Sud - B\^{a}t 425\\
F-91405 Orsay Cedex\\ France.
}
\email{bernard.helffer@math.u-psud.fr}

\date{\today}

\begin{document}

\bibliographystyle{plain}

\begin{abstract}
We consider the Neumann Laplacian with constant magnetic field on a regular domain. 
Let $B$ be the strength of the magnetic field, and let $\lambda_1(B)$ be the first eigenvalue of the magnetic Neumann Laplacian on the domain.
It is proved that $B \mapsto \lambda_1(B)$ is monotone increasing for large $B$.
Combined with the results of \cite{FournaisHelffer3}, this implies that 
all the `third' critical fields for strongly Type II
superconductors coincide.
\end{abstract} 

\maketitle

\section{Introduction and main result}
Let $\Omega \subset {\mathbb R}^2$ be a bounded, simply connected domain with regular boundary.
We keep this assumption in the entire paper.\\ 
Let ${\bf F}(x) = (F_1,F_2) = (-x_2/2, x_1/2)$---a standard choice for a vector potential generating a unit magnetic field: $\curl {\bf F} = 1$.
We consider ${\mathcal H}(B)$, the self-adjoint operator associated with the closed, symmetric quadratic form,
\begin{align*}
W^{1,2}(\Omega) \ni u \mapsto Q_B(u)=\int_{\Omega} |(-i\nabla + B {\bf F})u|^2\,dx.
\end{align*}
We will use the notation $p_{{\bf A}} = (-i\nabla + {\bf A})$. Then, 
more explicitly, ${\mathcal H}(B)$ is the differential operator $p_{B {\bf F}}^2$ with domain $\{ u \in W^{2,2}(\Omega) \, \Big|\, \nu \cdot p_{B {\bf F}}u |_{\partial \Omega} = 0 \}$, where $\nu$ is the unit interior normal to $\partial \Omega$.

We choose and fix a smooth parametrization $\gamma:  \frac{|\partial \Omega|}{2 \pi} {\mathbb S}^1\mapsto \partial \Omega$ of the boundary. We may assume that $|\gamma'(s)|=1$ for all $s$. We will further parametrize $\frac{|\partial \Omega|}{2 \pi}{\mathbb S}^1$ by $[-|\partial \Omega|/2, |\partial \Omega|/2]$ with periodicity being tacitly understood.

For a point $p = \gamma(s) \in \partial \Omega$ we define $k(p)$---also denoted by $k(s)$---to be the curvature of the boundary at the point $\gamma(s)$, i.e.
$$
\gamma''(s) = k(s) \nu(s),
$$
where $\nu(s)$ is the interior normal (to the boundary) vector at the point $\gamma(s)$. The maximum of $k$ will play an important role, we define therefore,
$k_{\rm max} := \max_s \{ k(s) \}$.

Define $\lambda_1(B)= \inf \Spec {\mathcal H}(B)$ to be the lowest eigenvalue of ${\mathcal H}(B)$. The diamagnetic inequality tells us that
$$
\lambda_1(B) \geq \lambda_1(0),
$$
for all $B\geq 0$.\\
 One may ask whether the more general inequality
$$
0< B_1 < B_2 \quad \Rightarrow \quad \lambda_1(B_1) \leq \lambda_1(B_2),
$$
which one can consider as a strong form of diamagnetism, holds (see \cite{Erd1},
 \cite{Erd2} 
 and \cite{LoTha}). 

In this paper we prove that strong diamagnetism holds for sufficiently large $B$.
\begin{thm}\label{thm:Derivative}~\\
The one sided derivatives,
$$
\lambda_{1,+}'(B) = \lim_{\epsilon \rightarrow 0_{+}} \frac{\lambda_1(B+\epsilon) - \lambda_1(B)}{\epsilon},
\quad\quad\lambda_{1,-}'(B) = \lim_{\epsilon \rightarrow 0_{+}} \frac{\lambda_1(B) - \lambda_1(B-\epsilon)}{\epsilon}
$$
exist for all $B>0$ and $\lambda_{1,+}'(B)$ satisfies
\begin{align}
\label{eq:LimitDerivative}
\liminf_{B \rightarrow \infty} \lambda_{1,+}'(B)  > 0.
\end{align}
Furthermore, there exists a universal constant $\Theta_0 > 0$ such that if $\Omega$ is not a disc, then the limit actually exists and satisfies,
\begin{align}
\label{eq:PreciseNotDisc}
\lim_{B \rightarrow \infty} \lambda_{1,-}'(B) =
\lim_{B \rightarrow \infty} \lambda_{1,+}'(B) = \Theta_0.
\end{align}
If $\Omega$ is a disc, then 
\begin{align*}
&\limsup_{B \rightarrow \infty} \lambda_{1,+}'(B)  >  \Theta_0, \\
&0 < \liminf_{B \rightarrow \infty} \lambda_{1,+}'(B)  < \Theta_0.
\end{align*}
In particular, in any case, there exists $B_0>0$ such that $\lambda_1(B)$ is strictly increasing on $[B_0, \infty)$.
\end{thm}

Results similar to \eqref{eq:LimitDerivative} have been proved recently in \cite{FournaisHelffer3} under extra assumptions. First of all (in \cite{FournaisHelffer2}) a complete asymptotics of $\lambda_1(B)$ was derived for $\Omega$ satisfying a certain `generic' assumption, i.e. that the boundary curvature only has a finite number of maxima, all being non-degenerate. This complete asymptotics was then used to obtain \eqref{eq:LimitDerivative}. 
The most prominent domain excluded in this approach is the disc---where the curvature is constant. However, \cite{FournaisHelffer3} includes a special analysis of the disc proving that Theorem~\ref{thm:Derivative} remains true in that case.

What remained was the study of all the other `non-generic' cases. Also it seemed desirable to be able to establish Theorem~\ref{thm:Derivative} 
without  using the existence of a complete asymptotic expansion, since such expansions are difficult to obtain and their structure depends heavily on the different kinds of maxima of the boundary curvature. In this paper we realize such a strategy. It turns out that for all domains, {\it except the disc}, one can modify the approach from \cite{FournaisHelffer3} to obtain \eqref{eq:LimitDerivative} with only very limited knowledge on the asymptotic behavior of $\lambda_1(B)$. For the disc one can use the special symmetry (separation of variables) of the domain to conclude.

Thus the structure of the proof of Theorem~\ref{thm:Derivative} is as follows. The statements for the disc follow from the analysis in \cite{FournaisHelffer3} which will not be repeated. 
Thus we only consider the case where $\Omega$ is not a disc.
If $\Omega$ is not a disc then there exists a part of the boundary where the ground state will be very small. Thus one can choose a gauge such that $|\widehat {\bf A} \psi| \ll 1$ (for large $B$ and in the $L^2$-sense),  where $\widehat {\bf A}$ is the vector field ${\bf F}$ in the new gauge. This new input to the proof in \cite{FournaisHelffer3} allows us to differentiate the leading order asymptotics for $\lambda_1(B)$.

Notice that if $\Omega$ is not a disc, then it satisfies the following assumption~:
\begin{assumption}\label{assump:Notdisc}~\\
If we  denote by $\Pi$ the set of maxima for the curvature, i.e.
$$
\Pi = \{ p \in \partial \Omega \,\big | \,k(p) = k_{\rm max} \},
$$
then $$
\Pi \neq \pa \Omega\;.
$$
\end{assumption}
Finally, we will prove  in Section \ref{s3} (Theorem \ref{thm:HPimproved}) 
 that all the natural definitions    of 
 the  third  critical field appearing in the theory of superconductivity coincide
without any other geometric assumption than regularity and simply connectedness.\\

\section{The analysis of the diamagnetism}
Two universal constants $\Theta_0, C_1$ will play an important role in this paper, as in any investigation of the magnetic Neumann Laplacian. For detailed information about these constants, one can refer to \cite{He-Mo}. 
For the second constant $C_1$,  we only use the fact that it is strictly positive.
The first, $\Theta_0$ can be defined as the ground state energy of the magnetic Neumann Laplacian with unit magnetic field in the case of the half-plane,
$$
\Theta_0 := \lambda_1(B=1), \quad\quad \text{ for } \quad\quad \Omega = {\mathbb R}^2_{+}.
$$
The numerical value of $\Theta_0$ can be calculated with precision ($\Theta_0 \approx 0. 59$), however for our purposes the following (easily established) rigorous bounds
$$
0< \Theta_0 < 1,
$$
suffice. 

We recall the following general, leading order asymptotics of $\lambda_1(B)$ proved in \cite{He-Mo}.

\begin{prop}~\\
As $B \rightarrow + \infty$, then 
\begin{align}
\label{eq:leading}
\lambda_1(B) = \Theta_0 B + o(B)\,.
\end{align}
\end{prop}

If a state $u$ is localized away from the boundary, i.e. $u \in C_0^{\infty}(\Omega)$, we have the following standard inequality
$$
\langle u \,,\, {\mathcal H}(B) u \rangle \geq B \| u \|_{L^2(\Omega)}^2\,,
$$
where, for $v$, $w$ in $L^2(\Omega)$, 
  $\langle v\,,\,w\rangle$ denotes the $L^2$ scalar product of $v$ and $w$.\\
Using that $\Theta_0<1$ it is therefore a standard consequence of \eqref{eq:leading} (for the proof see \cite{He-Mo}) that ground states are exponentially localized near the boundary.

\begin{lemma}[Normal Agmon estimates]\label{lem:NormalAgmon}~\\
There exists $\alpha, M,C>0$ such that if $B\geq 1$ and $\psi_1(\,\cdot\,;B)$ is a ground state of ${\mathcal H}(B)$ then
\begin{align}
\int_{\Omega} e^{2\alpha \sqrt{B} \dist(x,\partial \Omega)} &\big\{
|\psi_1(x;B) |^2 + \frac{1}{B} | p_{B{\bf F}} \psi_1(\,\cdot\,;B) |^2 \big\}\,dx\nonumber\\
&\leq C \int_{\{ \sqrt{B} \dist(x,\partial \Omega) \leq M\}}|\psi_1(x;B) |^2
 \,dx\, .
\end{align}
In particular, for all $N>0$,
\begin{align}
\int \dist(x,\partial \Omega)^N |\psi_1(x;B) |^2 \,dx = {\mathcal O}(B^{-N/2}).
\end{align}
\end{lemma}

>From \cite[Proposition 10.5]{He-Mo} we also get the following (stronger than \eqref{eq:leading}) result,

\begin{prop}\label{prop:LowerPotential}~\\
Let $\Theta_0, C_1$ be the usual universal constants and define, for $C>0$
$$
U_B(x) = \begin{cases} B, & \dist(x,\partial \Omega) \geq 2 B^{-1/6},\\
\Theta_0 B - C_1 k(s)\sqrt{B} - C B^{1/3},& \dist(x,\partial \Omega) \leq 2 B^{-1/6}.
\end{cases}
$$
Then, if $B\geq 1$ and $C$ is sufficiently big, we have for all $\psi \in W^{2,2}(\Omega)$,
$$
\langle \psi \,,\, {\mathcal H}(B) \psi \rangle
\geq \int_{\Omega} U_B(x) |\psi(x)|^2\,dx.
$$
\end{prop}

Proposition~\ref{prop:LowerPotential} and a corresponding improved upper bound (also proved in \cite{He-Mo}),
\begin{align}
\label{eq:BetterAsymp}
\lambda_1(B) = \Theta_0 B - C_1 k_{\rm max} \sqrt{B} + o(\sqrt{B}),
\end{align}
imply, by suitable Agmon estimates,  that ground states have to be localized near the set $\Pi$. We actually only need the following very weak version of this localization.
\begin{lemma}\label{lem:localisation}~\\
Let $\epsilon_0 >0$. Then, for all $N>0$, there  exists $C>0$ such that if 
$\psi_1(\,\cdot\,;B)$ is a ground state for ${\mathcal H}(B)$, then
$$
\int_{\{\dist(x,\Pi) \geq \epsilon_0\}} |\psi_1(x;B) |^2 \,dx\leq C \; B^{-N}\;.
$$
\end{lemma}

We now introduce adapted coordinates near the boundary. Define, for $t_0>0$
\begin{align*}
& \Phi : \frac{|\partial \Omega|}{2 \pi} {\mathbb S}^1\times (0,t_0) \rightarrow \Omega
& \Phi(s,t) = \gamma(s) + t \nu(s).
\end{align*}
For $t_0$ sufficiently small we have that $\dist(\Phi(s,t), \partial \Omega) = t$ and that
$\Phi$ is a diffeomorphism with image $\{ x \in \Omega \,| \, \dist(x, \partial \Omega) < t_0 \}$.
Furthermore, the Jacobian satisfies $|D\Phi| = 1-tk(s)$.

\begin{lemma}\label{lem:gauge}~\\
Let us  define for $\epsilon \leq  \min( t_0/2, |\partial \Omega|/2)$ and $s_0\in \pa \Omega$
$$
\Omega(\epsilon,s_0) := \{ x = \Phi(s,t) \, \big | \, t \leq \epsilon, |s-s_0| \geq \epsilon \}.
$$
Then there exists $\phi \in C^{\infty}(\Omega)$ such that
$\widehat {\bf A} = {\bf F}+\nabla \phi$ satisfies
$$
|\widehat  {\bf A}(x) | \leq C\, \dist(x, \partial \Omega),
$$
for $x \in \Omega(\epsilon,s_0)$.
\end{lemma}

\begin{proof}~\\
Let $\widetilde{\bf A}=(\widetilde{A}_1, \widetilde{A}_2)$ be the magnetic $1$-form pulled back to $(s,t)$ coordinates,
$$
F_1 dx + F_2 dy = \widetilde{A}_1ds+  \widetilde{A}_2 dt.
$$
Taking the exterior derivative, and using $dx \wedge dy = |D\Phi| ds \wedge dt$, we find
$$
\curl_{s,t}\widetilde{\bf A} =  \partial_s  \widetilde{A}_2 - \partial_t  \widetilde{A}_1 = (1-tk(s)).
$$
Since $\{(s,t) \, | \, t \leq \epsilon, |s-s_0| \geq \epsilon \}$ is simply connected there exists a function $\widetilde{\phi} \in C^{\infty}(\Phi^{-1}(\Omega(\epsilon,s_0)))$ such that
$$
\widetilde{\bf A} + \nabla_{s,t} \widetilde{\phi} = (t - t^2k(s)/2,0).
$$
Let $\chi \in C^{\infty}(\overline{\Omega})$,
\begin{align*}
&\chi = 1 \quad\text{ on } \quad \{x  \, | \, t \leq \epsilon, |s-s_0| \geq \epsilon \}, \\
&\chi =0 \quad\text{ on } \quad \{x  \, | \, \dist(x, \partial \Omega) \geq 2 \epsilon \mbox{ or }  |s-s_0| \leq \epsilon/2 \} ,
\end{align*}
and define $\phi(x) = \widetilde{\phi}(\Phi^{-1}(x)) \chi(x)$. Then $\phi$ solves the problem.
\end{proof}

\begin{proof}[Proof of Theorem~\ref{thm:Derivative}] 
$\,$\\
Let $\overline\phi \in C^{\infty}(\overline{\Omega})$ be such that ${\bf \overline F}:= {\bf F} + \nabla \overline\phi$ satisfies ${\bf \overline F} \cdot \nu = 0$ on $\partial \Omega$. The existence of such a $ \overline \phi$ is easy to prove. 
Define $\overline{ {\mathcal H}}(B)$ to be the self-adjoint operator associated to the closed quadratic form
$$
W^{1,2}(\Omega) \ni u\mapsto   \int_\Omega|-i\nabla u + B {\bf \overline F} u|^2 dx\,.
$$
Then $\overline{ {\mathcal H}}(B)$ and ${\mathcal H}(B)$ are unitarily equivalent and so they have the same spectrum. Furthermore, the domain of $\overline{ {\mathcal H}}(B)$ is
$$
{\mathcal D}(\overline{ {\mathcal H}}(B)) = \{ u \in W^{2,2}(\Omega) \,:\, \nu \cdot \nabla u \,\big|_{\partial \Omega} = 0 \},
$$
in particular, ${\mathcal D}(\overline{ {\mathcal H}}(B))$ is independent of $B$.
Applying analytic perturbation theory to $\overline{ {\mathcal H}}(B)$ we get the 
existence of $\lambda_{1,+}'(B), \lambda_{1,-}'(B)$. 

We recall that Theorem~\ref{thm:Derivative} was proved already in \cite{FournaisHelffer2}
in the case of the disk, so it remains to  consider the case where $\Omega$ is not the disc. Thus $\Omega$ satisfies Assumption~\ref{assump:Notdisc}.
Therefore, there exist $s_0 \in [-|\partial \Omega|/2, |\partial \Omega|/2]$ and $0 < \epsilon_0 < \min( t_0/2, |\partial \Omega|/4)$ such that
$$
[s_0 - 2\epsilon_0, s_0 + 2\epsilon_0] \cap \Pi = \emptyset.
$$
Let $\widehat {\bf A} $ be the vector potential defined in
Lemma~\ref{lem:gauge},  $\widehat Q_B$ the quadratic form
$$
W^{1,2}(\Omega) \ni u\mapsto
\widehat Q_B(u)=  \int_\Omega|-i\nabla u + B \widehat {\bf A}u|^2 dx\,,
$$
and
$\widehat {\mathcal H}(B)$ be the associated operator. 
Then $\widehat {\mathcal H}(B)$ and $\overline{\mathcal H}(B)$ are unitarily
equivalent: $\widehat {\mathcal H}(B)=e^{iB\phi}\overline{\mathcal H}(B)e^{-iB\phi}$, for some $\phi$ independent of $B$.
By analytic perturbation theory applied to $\overline{\mathcal H}(B)$ there exists an analytic branch of eigenfunctions,
$$
\overline{\mathcal H}(\beta) \overline{\psi}_{1,+}(\,\cdot\,;\beta) = \lambda_1(\beta) \overline{\psi}_{1,+}(\,\cdot\,;\beta),
$$
for $\beta \in [B, B+\epsilon)$, some $\epsilon>0$, with $\| \overline{\psi}_{1,+}(\beta) \| = 1$.\\
With $\psi^+_1(\,\cdot\,;\beta):=e^{i\beta\phi} \overline{\psi}_{1,+}(\,\cdot\,;\beta)$ being the corresponding eigenfunctions of
$\widehat{\mathcal H}(\beta)$, we get
\begin{align}
\label{eq:FormDeriv}
\lambda_{1,+}'(B) &= \frac{d}{d\beta} \widehat Q_{\beta}(\psi^+_1(\beta)) \big|_{\beta=B}\nonumber\\
&= \langle \widehat {\bf A} \psi^+_1(\,\cdot\,;B) \,,\,  p_{B 
\widehat {\bf A}}
\psi^+_1(\,\cdot\,;B)\rangle +  \langle  p_{B 
\widehat {\bf A}}  \psi^+_1(\,\cdot\,;B) \,,\, \widehat {\bf A}
\psi^+_1(\,\cdot\,;B)\rangle \nonumber\\
&\quad+ 2 \Re \{ \widehat Q_B(v, \psi^+_1(B)) \}, 
\end{align}
where $v = \frac{d}{d\beta} \psi^+_1(\beta) \big|_{\beta=B}$.
The last term in \eqref{eq:FormDeriv} vanishes because $\psi^+_1$ is a normalized eigenfunction of $\widehat {\mathcal H}$, and therefore,
\begin{equation}
\lambda_{1,+}'(B) = \langle \widehat {\bf A} \psi^+_1(\,\cdot\,;B) \,,\,  p_{B 
\widehat {\bf A}}
\psi^+_1(\,\cdot\,;B)\rangle +  \langle  p_{B 
\widehat {\bf A}}  \psi^+_1(\,\cdot\,;B) \,,\, \widehat {\bf A}
\psi^+_1(\,\cdot\,;B)\rangle \;.
\end{equation}
We now obtain for any $\beta >0$, 
\begin{align}
\lambda_{1,+}'(B) & =  \frac{\widehat{Q}_{B+\beta}(\psi^+_1(\,\cdot\,;B)) - \widehat{Q}_B (\psi^+_1(\,\cdot\,;B))}{\beta}  - \beta \int_{\Omega} \vert \widehat {\bf A} \vert ^2 \, |\psi^+_1(x;B)|^2\,dx \nonumber\\
&\geq \frac{\lambda_1(B+\beta) - \lambda_1(B)}{\beta} - \beta \int_{\Omega} \vert \widehat {\bf A} \vert ^2 \, |\psi^+_1(x;B)|^2\,dx\,.
\end{align}
By Lemma~\ref{lem:gauge} we can estimate
\begin{align}
\int_{\Omega} \vert \widehat {\bf A} \vert ^2\; |\psi^+_1(x;B)|^2\,dx
&\leq C \int_{\Omega} \dist(x,\partial \Omega)^2 |\psi^+_1(x;B)|^2\,dx \nonumber\\
&\quad+
\|\widehat  {\bf A}  \|_{\infty}^2 \int_{\Omega \setminus \Omega(\epsilon_0,s_0)} |\psi^+_1(x;B)|^2\,dx.
\end{align}
Combining Lemmas~\ref{lem:NormalAgmon} and ~\ref{lem:localisation} we therefore find the existence of a constant $C>0$ such that~:
\begin{align}
\int_{\Omega} \vert \widehat {\bf A} \vert ^2\, |\psi^+_1(x;B)|^2\,dx \leq C\, B^{-1}\,.
\end{align}
We now choose $\beta = \eta\, B$, where $\eta>0$ is arbitrary. By the weak asymptotics \eqref{eq:leading} for $\lambda_1(B)$, we therefore find~:
\begin{align}
\liminf_{B \rightarrow \infty} \lambda_{1,+}'(B) \geq \Theta_0 - \eta\, C\,.
\end{align}
Since $\eta$ was arbitrary this implies
\begin{align}
\liminf_{B \rightarrow \infty} \lambda_{1,+}'(B) \geq \Theta_0\,.
\end{align}
Applying the same argument to the derivative from the left, $\lambda_{1,-}'(B)$, we get (the inequality gets turned since $\beta<0$)
\begin{align}
\limsup_{B \rightarrow \infty} \lambda_{1,-}'(B) \leq \Theta_0.
\end{align}
Since, by perturbation theory, $\lambda_{1,+}'(B) \leq \lambda_{1,-}'(B)$ for all $B$, we get \eqref{eq:PreciseNotDisc}.
\end{proof}
\section{Application to superconductivity}\label{s3}
 As appeared 
 from  the works of Bernoff-Sternberg~\cite{BeSt},  Del Pino-Felmer-Stern\-berg \cite{PiFeSt}, Lu-Pan~\cite{LuPa1, LuPa2, LuPa3}, and Helffer-Pan~\cite{He-Pan}, the determination of 
the lowest eigenvalues of the magnetic Schr\"{o}dinger operator is
crucial for a detailed description of the nucleation of
superconductivity (on the boundary) for superconductors of Type II and
for accurate estimates of the critical field $H_{C_3}$. 
In this section we will clarify the relation between the different definitions
of critical fields considered in the mathematical or physical literature
and all supposed to describe the same quantity.
This is a continuation and an improvement of \cite{FournaisHelffer3}~:
we will be indeed able to eliminate all the geometric assumptions of that paper.\\

We recall that the Ginzburg-Landau functional is given by
\begin{multline}
\label{eq:GL_F}
{\mathcal E}[\psi,{\bf A}] = {\mathcal
E}_{\kappa,H}[\psi,{\bf A}]  =
\int_{\Omega} \Big\{ |p_{\kappa H {\bf A}}\psi|^2 
- \kappa^2|\psi|^2
+\frac{\kappa^2}{2}|\psi|^4 \\
+ \kappa^2 H^2
|\curl {\bf A} - 1|^2\Big\}\,dx\;,
\end{multline}
with
$ (\psi, {\bf A})  \in W^{1,2}(\Omega;{\mathbb C})\times
W^{1,2}(\Omega;{\mathbb R}^2)$. 

We fix the choice of gauge by imposing that 
\begin{align}
\label{eq:gauge}
\Div {\bf A} &= 0 \quad \text{ in } \Omega\;, & {\bf A} \cdot \nu = 0 \quad \text{ on } \partial \Omega\;.
\end{align}
We recall that the domains $\Omega$ are assumed to  be smooth, bounded and simply-connected and  refer the reader to \cite{Bo2},\cite{BoDa} and \cite{BoFo} for the analysis of the case with corners.

By variation around a minimum for ${\mathcal E}_{\kappa,H}$ we find that minimizers $(\psi, {\bf A})$ satisfy the Ginzburg-Landau equations,
\begin{subequations}
\label{eq:GL}
\begin{align}
\left.\begin{array}{c}
p_{\kappa H {\bf A}}^2\psi =
\kappa^2(1-|\psi|^2)\psi \\
\label{eq:equationA}
\curl^2 {\bf A} =-\tfrac{i}{2\kappa H}(\overline{\psi} \nabla
\psi - \psi \nabla \overline{\psi}) - |\psi|^2 {\bf A}
\end{array}\right\} &\quad \text{ in } \quad \Omega \, ;\\
\left. \begin{array}{c}
(p_{\kappa H {\bf A}} \psi) \cdot \nu = 0 \\
\curl {\bf A} - 1 = 0
\end{array} \right\} &\quad \text{ on } \quad \partial\Omega \, ,
\end{align}
\end{subequations}
with
$$
\curl^2 {\bf A} =
(\partial_{x_2}(\curl {\bf A}),-\partial_{x_1}(\curl {\bf
A})) \, .
$$

It is known that, for given values of the parameters $\kappa, H$, the functional ${\mathcal E}$ has (possibly non-unique) minimizers. However, after some analysis of the functional, one finds (see \cite{Giorgi-Phillips} for details)
 that, for any  $\kappa >0$,  there exists $H(\kappa)$ such that if $H>H(\kappa)$ then $(0,{\bf F}_\Omega)$ is the only minimizer of ${\mathcal E}_{\kappa,H}$ (up to change of gauge).\\
Here we choose ${\bf F}_\Omega$ as the unique solution in $\Omega$
 of
$\curl {\bf F}_\Omega =1$ satisfying \eqref{eq:gauge}.
Following Lu and Pan \cite{LuPa1}, one  can therefore first define
\begin{align}
\underline{H}_{C_3}(\kappa) = \inf\{ H>0 \;:\; (0, {\bf F}_\Omega) \text{ is a minimizer of } {\mathcal E}_{\kappa,H}\}\;.
\end{align}
In the physical interpretation of a minimizer $(\psi,{\bf A})$, $|\psi(x)|$ is a measure of the superconducting properties of the material near the point $x$. Therefore, $\underline{H}_{C_3}(\kappa)$ is the value of the external magnetic field, $H$, at which the material loses its superconductivity completely.

Actually, as already used implicitly in \cite{LuPa1}
 and more explicitly in \cite{FournaisHelffer3}, we should also introduce   an upper critical field, 
$\underline{H}_{C_3}(\kappa)\leq \overline{H}_{C_3}(\kappa)$, by
\begin{align}
\overline{H}_{C_3}(\kappa) &=  \inf\{ H>0 \;:\; \text{for all } H'>H,
 (0, {\bf F}_\Omega) \text{ is the only minimizer of } {\mathcal E}_{\kappa,H'}\} \;.
\end{align}
$\,$

The physical idea of a sharp transition from the superconducting to the normal state, requires the different 
definitions of the critical field to coincide.

Most works analyzing $\underline{H}_{C_3}$ relate (more or less implicitly)
 these {\it global} critical fields to {\it local} ones given purely 
in terms of spectral data of the  magnetic Schr\"{o}dinger operator ${\mathcal H}(B)$, 
 i.e. in terms of a {\it linear} problem. The local fields are defined as follows.
\begin{align}
\label{eq:SpecLocalDef}
\overline{H}_{C_3}^{\rm loc}(\kappa) &=  \inf\{ H>0 \;:\; \text{ for all } H'>H, \quad \lambda_1(\kappa H') \geq \kappa^2 \} \;, \nonumber \\
\underline{H}_{C_3}^{\rm loc}(\kappa) &=  \inf\{ H>0 \;:\;  \lambda_1(\kappa H) \geq \kappa^2 \}\;.
\end{align}

The difference between $\overline{H}_{C_3}^{\rm loc}(\kappa)$ and
 $\underline{H}_{C_3}^{\rm loc}(\kappa)$---and also between 
$\overline{H}_{C_3}(\kappa)$ and $\underline{H}_{C_3}(\kappa)$---can be retraced to the general non-existence of an inverse to the function $B \mapsto \lambda_1(B)$, i.e. to lack of strict monotonicity of $\lambda_1$.
In the previous section, we have solved  this monotonicity question
and we now explain, following mainly  \cite{FournaisHelffer3}, how
this permits to close the discussion about this `third' critical field
in the high $\kappa$ regime.

The next theorem, which  is proved in \cite{FournaisHelffer3}, is typical of Type II materials, in the sense that it is only valid for large values of $\kappa$.

\begin{thm}
\label{thm:Identical}~\\
There exists a constant $\kappa_0>0$ such that, for $\kappa > \kappa_0$,  we have
\begin{align}
\underline{H}_{C_3}(\kappa) = \underline{H}_{C_3}^{{\rm loc}}(\kappa)\;,
\quad\quad
\overline{H}_{C_3}(\kappa) = \overline{H}_{C_3}^{{\rm loc}}(\kappa)\;.
\end{align}
\end{thm}

$\,$

On the other hand,  we have from Theorem \ref{thm:Derivative}~:
\begin{prop}
\label{prop:InverseFunction}~\\
There exists $\kappa_0$ such that, if $\kappa \geq \kappa_0$, then the equation for $H$:
\begin{align}
\label{eq:FormalField}
\lambda_1(\kappa H) = \kappa^2\;,
\end{align}
has a unique solution $H(\kappa)$. 
\end{prop}
In other words, for large $\kappa$, the upper and lower {\it local} fields, defined in \eqref{eq:SpecLocalDef}, coincide.
We define, for $\kappa \geq \kappa_0$, the local critical field $H_{C_3}^{{\rm loc}}(\kappa)$ to be the solution given by Proposition~\ref{prop:InverseFunction}, i.e.
\begin{align}
\label{eq:Hnobar}
\lambda_1(\kappa H_{C_3}^{{\rm loc}}(\kappa)) = \kappa^2\;.
\end{align}

Using Proposition~\ref{prop:InverseFunction} we can identify the lower and upper local fields and therefore find the following result.

\begin{thm}
\label{thm:HPimproved}~\\
Suppose $\Omega$ is smooth, bounded  and simply connected. 
There exists $\kappa_0 >0$ such that, when $\kappa>\kappa_0$, then
\begin{align}
H_{C_3}^{{\rm loc}}(\kappa) = \underline{H}_{C_3}(\kappa) =
\overline{H}_{C_3}(\kappa)\;.
\end{align}
\end{thm}

\begin{remark}~\\
This result was established in \cite{FournaisHelffer3} under the additional
 assumption that $\Omega$ was either a disk or a domain whose boundary
 has only a finite number of points of maximal curvature (with in addition
 some non degeneracy condition).
\end{remark}

\end{document}